\documentstyle[aps,prl,twocolumn,epsf]{revtex}
\begin{document}
\twocolumn[\hsize\textwidth\columnwidth\hsize\csname@twocolumnfalse\endcsname
\draft
\preprint{}
\title{Quantum computation by optically coupled steady atoms/quantum-dots inside
       a  quantum electro-dynamic cavity}

\author{Prabhakar Pradhan$^1$\cite{ppemail},  M.P. Anantram$^2$ and
    Kang L. Wang$^1$}
\address{$^1$Electrical Engineering Department, University of California,
          Los Angeles, CA 90095 \\
         $^2$Nasa Ames Research Center, Moffett Field, CA 94035}
\maketitle
\date{\today}
\begin{abstract}
We present a model for quantum computation using $n$ steady 3-level 
atoms or 3-level quantum dots, kept inside a quantum electro-dynamics
(QED) cavity. Our model allows one-qubit operations and the two-qubit
controlled-NOT gate as required for universal quantum computation.
The $n$ quantum bits are described by two energy levels of each atom/dot.
An external laser and $n$ separate pairs of electrodes are used to
address a single atom/dot independent of the others, via Stark effect.
The third level of each system and an additional common-mode qubit
(a cavity photon) are used for realizing the controlled-NOT operation
between any pair of qubits. Laser frequency, cavity frequency, and 
energy levels are far off-resonance, and they are brought to resonance
by modifying the energy-levels of a 3-level system using the Stark 
effect, only at the time of operation.
\end{abstract}
\pacs{03.67.Lx, 73.20.Dx, 42.50.Md}
]
A computer, which follows
quantum  mechanical
principles, has significant advantages over a classical 
computer~\cite{reviews,shor}.
Implementing a quantum computer is based upon the implementation 
of basic quantum units called quantum bits (two-level systems)
and  communication among them.
Logical operations of a
quantum computer can be decomposed into a series of
 an arbitrary one-qubit rotation
plus a two-qubit controlled-NOT
operation, thus this set of operations makes a universal
quantum computer \cite{sw,9co}.
A similar set in which the controlled-NOT is replaced by the 
controlled-phase-shift gate [$|00\rangle \rightarrow |00\rangle$,
$|01\rangle \rightarrow |01\rangle$,
$|10\rangle \rightarrow |10\rangle$,
and $|11\rangle \rightarrow -|11\rangle$] is also universal~\cite{cz}.
$n$ two-level systems can have $2^n$ highly entangled
(phase coherence) states and a quantum
computer takes  advantages of  performing  unitary transformations
in a parallel manner on these $2^n$ classical strings.

The main difficulties in implementing a quantum computer are the
contradicting demands in terms of interaction with the environment.
On one hand, a strong {\em controlled} interaction is desired in order
to operate the computing algorithm (to switch the state of the qubits),
but on the other hand {\em uncontrolled} interactions are strongly
undesired since they cause decoherence of the qubits and hence loss
of computing ability.
All  quantum systems lose their coherence after some time due to 
non-zero coupling with the environment. Thus the above problem is 
usually expressed as the need to increase the ratio between the 
decoherence time and the switching time: a quantum computer must 
perform {\em all} calculations within the decoherence time of the 
qubit.

Several theoretical and experimental attempts are currently 
ongoing  to realize simple gates (such as the controlled-NOT gate
between two qubits). The most realistic ones at the moment are
ion-trap \cite{cz}, liquid NMR~\cite{nmr}, 
and cavity-QED \cite{sw,haroche,kimble}.
However, serious problems in scaling these systems, and/or in 
addressing particular qubits create the need for better suggestions or
major modifications of these implementations.

The first interesting experiments were done on cavity-QED
systems \cite{kimble}.
The cavity-QED computation model is based on the
idea of having two types of qubits (atoms and cavity-modes)
and it was found very useful in implementing various gates.
However, the requirement of mechanical control of atoms makes this
model less desirable for quantum computing: the interaction 
time is controlled by the physical motion of the atoms inside the cavities,
and having enough control to let the atom enter a cavity several times
(whenever required by the algorithm) is difficult.

The long decoherence time of  liquid-NMR  and ion trap systems,
and easy control of the qubits make these systems a  serious         
candidate for quantum computation. But the problem is in
scaling these systems with the increasing numbers of qubits.
Like, in case of bulk liquid-NMR, the signal from the system
decreases exponentially with the numbers of qubits.
Thus the number of qubits comprising the quantum computer has to be
small.  The ion-trap computation model is based on interaction via a 
common-mode qubit and it has two main problems:
(a) the addressing of an individual qubit by a separate laser 
directed to each ion is an idea which cannot be implemented yet.
(b) the use of only one type of two-qubit interaction---interaction 
via one common mode.  This problem prevents the possibility of running
simultaneous several gate operations.

Proposals for a solid state~\cite{dl,gs} and solid-state 
NMR~\cite{kane} devices based on nanotechnology, might
be more promising for the far future. But even single qubit systems 
have not been implemented in these systems due to the difficulties 
in creating and controlling such a single qubit.

Clearly, more candidates for realization of quantum computing devices are
still needed, with the hope of a more diverted experimental effort.
Such an effort, mainly in the direction of solid-state devices,
but combining ideas from existing implementations, might lead to a system
where single qubits can be addressed, scaling to large number of qubits
made possible, and in the future, may be even a system where 
fault-tolerant computation can be performed. 

This paper suggests a model of a quantum computer that combines the
advantages of other models \cite{pp}.
We shall show how to implement the universal set of gates containing
the controlled-phase-shift and the arbitrary one qubit rotation.

As a first step we suggest a hypothetical model of atoms fixed in 
a cavity, and a pair of electrodes ``directed'' around each atom to 
control its energy level-spacing. 
In this first step, we combine the use of a common mode as in the 
ion-trap computation model \cite{cz}, with the two types of qubits as
suggested by cavity-QED models \cite{sw,haroche,kimble}, to obtain
better control of addressing a single qubit.
Unfortunately, fixing atoms for the required time scales is not yet 
realistic.
In general, atoms are in motion in all cavity QED experiments.

In the second step, we suggest replacing the atoms by quantum dots, 
so the idea of ``fixing'' the qubits becomes more realistic.
The technical ability of putting a single qubit in a single quantum
dot, and the technical ability of putting a quantum dot in a cavity 
exist separately.  Combining them together (while demanding also that
the cavity is highly reflecting) is far from the ability of current 
experiments, but we hope to motivate this direction by showing that 
the computation model we obtain is very promising.
Recently, it has been experimentally shown \cite{dot} that a single 
electron can be controlled in a quantum dot; The dot size and
dielectric modulation are however large $(.5 \mu m)^3$.

A sketch of the model for the proposed quantum-computer (with steady 
atoms) is shown in Fig.1.  Atoms are kept steady along the axis of the
cavity. An external laser source is accessible to all atoms, and is 
directed perpendicular to the cavity axis. Electrodes around each atom
(which we refer as ``Stark plates'') are used to control its level
spacing via the
Stark effect, and are perpendicular to the laser and the cavity axis.
When required in the protocol, a strong electric field is applied to the
atom by changing the voltage on the electrodes. This field changes the
energy level separation (a thorough study has been done for Rydberg 
atoms in Ref.\cite{mic,ran}). The required electric field can be 
calculated easily once the energy levels and the wave functions of the
system are chosen.
We will assume that the on/off switching of the electric field is
but slow such that the change in the original wave function is
insignificant. At the same time, it must be fast relative to the time
steps of the computation.
The applied DC field has to be a fraction
of the order of the atomic energy level separations.

Quantum bits of the computer are described by the ground state 
($|g\rangle$), and the first excited state ($|e_0\rangle$) of the atom;
a third level ($|e_1\rangle$) is used for a controlled phase shift 
operation.
Rotation of an individual qubit is
achieved by applying a laser pulse to all atoms, while only one the
qubit undergoing transformation is on-resonance with the laser 
frequency, and others are far off-resonance.

\begin{figure}
\epsfxsize=8cm
\centerline{\epsfbox{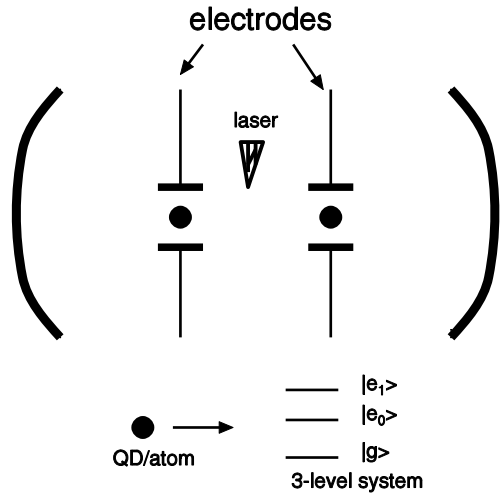}}
\vspace{.1cm}
\caption{ 
Atoms/QDs are kept along the axis of a perfectly reflecting cavity.
Electric plates are kept attached around each of the atoms/QDs to control 
the energy
levels via Stark effect. A laser source pointing toward
all atoms is kept perpendicular to  both, i.e., the cavity and the 
Stark plate's axes.}        
\label{pspace}
\end{figure}

Communication between any two qubits is done by a common mode cavity 
photon as described now. The photonic mode is in its ground state (zero 
photons) and a maximum of one cavity photon is present at the time of 
interaction between two qubits. The cavity's $0$-photon and $1$-photon
states are defined by $|0\rangle$ and $|1\rangle$ respectively.
The atomic levels are kept far off-resonance with respect to
the resonant frequency of the cavity, to avoid undesired interaction 
(in which a transition from excited state to
ground state takes place while emitting a photon to the cavity).
Desired energy levels are brought into resonance with the cavity by 
changing the Stark field only at the time of logic operations.
To perform a controlled operation between any two qubits,
we do the following:
(a) The state of the first qubit and the vacuum state
of the cavity are swapped. (b) the new cavity state is used to perform a 
controlled operation with another qubit;
the third level of the atom is
used for that purpose, yielding a controlled-phase-shift gate.
(c) Finally, the cavity state is
again swapped with the first qubit, so the cavity is back into its vacuum
state, and the controlled operation between the two qubits is completed.

We assume that the time for a significant  far off-resonance evolution
is huge compare to the on-resonance evolution time.
We also assume that the cavity is of high-quality and has almost
perfect reflecting walls, so that the decoherence time for the cavity mode
is much larger than the time between the two required swap operations.
The frequency of the laser pulses is off-resonant with the cavity.

We will describe in detail the Hamiltonian leading to the single qubit
rotations and the controlled-phase-shift operations. This is done by
taking into account the fact that $|g\rangle \rightarrow |e_0\rangle$
and $|e_0\rangle \rightarrow |e_1\rangle$ are allowed dipole
transitions, but $|g\rangle \rightarrow |e_1\rangle$ is not an allowed
transition due to the definite parity of the wave function.

Let $\omega_{g;e_0}$ be the level separations of the qubit (a similar
definition applies for $\omega_{e_0;e_1}$ and $\omega_{g;e_1}$).
For the atomic levels with definite parity,
we assume that the levels are chosen so that the difference frequencies
$\omega_{g;e_0}$ and $\omega_{e_0;e_1}$ are nearly the same.
We treat this case here, but one can easily treat the case where other
transitions are allowed or forbidden.

In the following, we describe the steps to obtain the necessary
operations involving only one atom at a time,
by bringing its levels to be on-resonance with the laser frequency or the
cavity mode.  
The other atoms are kept far off-resonance to avoid their interactions.
If the initial levels are such that $g;e_0$, and $e_0;e_1$ are the allowed
transitions, and $\omega_{e_0;g} < \omega_{e_0;e_1}$, then
one way to choose the cavity and the laser frequencies are such that
$\omega_{e_0;g} < \omega_{e_0;e_1} < \omega_l < \omega_c$. 

By increasing the level separations, the qubit can be brought to be 
on-resonance with the laser. This increase in level separation must be
significant enough that the interaction of off-resonant atoms with
the laser is insignificant. By increasing the level separations further,
the qubit is brought to be on-resonance with the cavity photon.
Each level separation increases with the applied electric
field. While increasing the level separation $\omega_{e_0;g}$, the level
separation $\omega_{e_0;e_1}$ will first come to resonance
with  the cavity. But we will assume that the switching
time is much smaller than the inverse Rabi frequency of the atom-cavity
system such that there is practically no effect of this resonance 
crossing.

{\em One qubit rotation} is performed by changing the atomic levels so
that $\omega_l = \omega_{e_0}-\omega_g$ and applying the laser pulse.
The laser and qubit involved interact on resonance (but $\omega_l$ is
off resonance with the cavity and with other qubits).
The Hamiltonian for the atomic levels in the presence of the
laser field is \cite{cz}:
\begin{eqnarray*}
  H_{1} = \frac{\Omega_l}{2} \; 
  [\sigma_+ e^{-i\phi} 
  +\sigma_- e^{i\phi} ] \mbox{.}  
\end{eqnarray*}
Where $\sigma_+=|e_0><g|$,  $\sigma_-=|g><e_0|$, 
$\phi$ is the phase factor
 of the laser at the location of 
the basic unit, and $\Omega_l$ is the Rabi frequency due to the laser
$\Omega_l=E_0\mu_{ge_0}$, where $\mu_{ge_0}$ is the  
dipole moment for $|g> \rightarrow |e_0>$ transition  and
$E_0$ is the strength of the electric field.

If the interaction time between the laser pulse and the qubit is
$t = \frac{k \pi}{\Omega_l}$, then the time evolution
operator is 
\begin{eqnarray*}
 \hat{V}_m^k(\phi) = exp[ -i k \frac{\pi}{2} (\sigma_+^m e^{-i\phi}
+ \sigma_-^m e^{i\phi}) ]
\end{eqnarray*} 
The process is an energy non-conserving process, and the system is fed
energy from the laser field.

The Jaynes-Cumming Hamiltonian \cite{haroche}  for a 2-level system,
which is on-resonance with the cavity photon is described by: 
\begin{eqnarray*}
H_{2} = i \frac{\Omega_c}{2} \; 
       [ \sigma_+\hat{a}^\dagger  - 
	\sigma_- \hat{a}  ] 
\end{eqnarray*}
Where $a$ and $a^\dagger$ are the annihilation and creation operators
for common mode photon, and
$\Omega_c$ is the  photon Rabi frequency of the cavity-atom system.

If the interaction time between the laser pulse and the qubit is 
$t = \frac{k \pi}{\Omega_c}$, then
 the time evolution operator is 
\begin{eqnarray*}
 \hat{U}_m^k(\phi) = exp[ -i k \frac{\pi}{2} 
    [ i\sigma_+^m \hat{a}^\dagger   
 -i\sigma_-^m \hat{a} ]] \mbox{.} 
\end{eqnarray*} 

To get a control-phase-shift between two qubits (two atoms/QDs, say $m$ and
$n$ such that $m$ is the control and $n$ is the target), 
we need two types of cavity-atom operations:
A $\pi$ pulse for obtaining the swap operation, where the qubit levels 
$g;e_0$, and the cavity levels are
used and a $2\pi$ pulse using the third level and the cavity photon 
to obtain the atom-cavity controlled-phase-shift.

The operation is done in three steps: \\
\indent (1)  The levels $|g\rangle_m$ and $|e_0\rangle_m$ of the $m$th atom
are brought into resonance with the cavity. The system is let to 
evolve on-resonance with the cavity for a time equal to $\pi/\Omega_c$.
At the end of this, a SWAP operation the state of the atom with the 
state of the cavity (which is the vacuum state) occurs.
After the interaction, the $m$th atom is in its ground state.

(2) The states $|g\rangle_n$ and $|e_1\rangle_n$ are brought to
resonance with the cavity and let to evolve for a time equal to
$2\pi/\Omega_c$.
The result is that the state doesn't change if there is no photon in the
cavity: $|g_n 0 \rangle \rightarrow |g0_n \rangle $,
$|e_{0n} 0 \rangle \rightarrow |e_{0n} 0 \rangle$;
Also, there is no change if the $n$th atom is in its ground state:
$|g 1 \rangle \rightarrow |g1 \rangle $; however, if there is a photon in
the cavity, and the $n$th atom is in the excited state, it gets a phase 
$|e_{0n} 1 \rangle \rightarrow -|e_{0n} 1 \rangle $ (since it is a spinor).

(3) The $m$th atom and the cavity are brought into resonance and the
system is let to evolve for a time equal to $\pi/\Omega_c$.
At the end of this, a SWAP operation the state of the $m$th atom with the 
state of the cavity (which is the vacuum state) occurs.

A $\pi$ pulse $(k=1)$ is given between the levels $|g\rangle_m$ and 
$|e_0\rangle_m$ by bringing these two levels (of the $m$'th atom)
on resonance with the cavity to SWAP again their states. After the 
interaction, the cavity is back in the vacuum state, but the state of
the qubits change.

A crucial issue in this model is the relative time scale between the
cavity  on-resonance and off-resonance with the 2-level system.
When the cavity is off-resonance in presence of a photon, a dressed
state evolves. The relative time scale of the 
evolution is \cite{haroche}:
\begin{eqnarray*}
\frac{\tau_{off-resonance}}{\tau_{on-resonance}} =
 \sqrt{( \frac{\omega_c-\omega_{ge_0}}{\Omega_c})^2 +1 } 
\end{eqnarray*}
where $\omega_c$ is the cavity frequency $\omega_{ge_0}=\omega_{e_0}-\omega_g$
and $\Omega_c$ is the
Rabi frequency of the atom due to the cavity photon. The vacuum off-resonance
phase evolution, $\Omega_c^2/(\omega_c-\omega_{e_0g})$ must be
small enough to make the off-resonant evolution insignificant
\cite{haroche}.
Another way to get rid of the vacuum off resonance evolution is by       
nullifying the extra  phase evolution by additional logic operations
or by taking into account the phase in every  step of operations.

The spontaneous emission time is quite low or negligible for a trapped
atom.
The ratio of decoherence time to the time required for a single
operation is $\approx10^6$. That is, $10^6$ pulses can be applied
within the coherence time.
In case of a Rydberg atom, for $50\rightarrow 51$ transition,
$\omega_{ge_0}=5*10^{10}$ Hz, the cavity length is $\approx 1$ cm,
$\Omega_c\approx 4*10^5$ Hz, and $\delta=4*10^6$ Hz, where $\delta$
is the detuning, i.e., $\omega_{e;g}-\omega_l$.

In the case of a quantum dot with transition energy $\omega_{ge_0}
\approx 1 meV$ (1 THz), $\Omega_c\approx 10^8$ Hz and the cavity length
is $\approx 150 \mu$. The phase coherence length should be much larger
than $1/\Omega_c$ to realize a system that is capable of performing non
trivial operations. The best reported values for decoherence times in a
quantum dot system are comparable to $1/\Omega_c$.~\cite{marcus} These
dots were however open in the sense that large electron reservoirs were
connected to them. Isolated quantum dots that are specifically designed
to reduce the decoherence times will be of paramount importance, not only
here but also in other applications of coherent phenomena.

Measurement of the final state of the quantum computer is crucial for 
an experiment. In this proposed model, qubits are inside the cavity, 
the  state of a qubit can be measured by the following procedure:
(a) Transferring the quantum state of the qubit to the cavity by 
bringing the qubit into resonance with the cavity and waiting for 
half the time period of the atom-cavity Rabi oscillation.
If the electron in the qubit is in higher state, it will release
a photon to the cavity. (b) This photon has to be detected from
the cavity by a detector, which is a difficulty that all models of 
quantum computing suffer from.

 Here we have shown a new model  of quantum-computer using
 atoms or quantum dots inside a quantum cavity.
 A similar model can be
 easily designed for spin-states inside a cavity by replacing the Stark
 effect by a Zeeman effect.
  With the advance of  technology, it may be possible to
  fabricate steady atoms inside the cavity or quantum dots inside a
  cavity with long enough decoherence time.
  The important point of this model is that the 
  qubits are easily addressed (and we don't require a separate laser
  addressing each one).
  Note that operations are done only when  the cavity/laser and
  the atomic levels are on-resonance, while undesired interactions are
  avoided by keeping the far
  off-resonance condition for the other atoms. 
 The main operations are done by an external laser and controlling
 the voltage of the stark plates from outside.

We are grateful to T. Mor and V. P. Roychowdhury for useful discussions 
throughout this work. The work of PP was  supported by 
the Revolutionary Computing group at Jet Propulsion
Laboratory, contract No. 961360, and grant No. 530-1415-01 from DARPA
Ultra program.

\end{document}